\documentclass{caosp302}
\usepackage{float}
\usepackage{graphicx}

\articleNo{123}
\pubyear{2007}
\volume{35}
\volnumber{3}
\firstpage{1}
\received{October, 2007}
\accepted{August 28, 2005}

\begin{document}

%
\htitle{Magnetic fields in massive stars}
\hauthor{S.\,Hubrig et al.}

\title{Magnetic fields in massive stars}


%
\author{
        S.~Hubrig\inst{1}
      \and 
        M.~Sch\"oller\inst{1}
        \and  
       M.~Briquet\inst{2}
       \and  
       M.A.~Pogodin\inst{3}
       \and 
       R.V.~Yudin\inst{3}
       \and
        J.F.~Gonz\'alez\inst{4}
       \and 
       T.~Morel\inst{2}
       \and 
      P.~De Cat\inst{5}
       \and 
        R.~Ignace\inst{6}
 \and 
        P.~North\inst{7}
 \and 
        G.~Mathys\inst{1}
 \and 
        G.J.~Peters\inst{8}
       }

%
\institute{
           ESO, Casilla 19001, Santiago 19, Chile \email{shubrig@eso.org}
         \and 
Instituut voor Sterrenkunde, Katholieke Universiteit Leuven,\\ Celestijnenlaan 200D, B-3001 Leuven, Belgium
         \and 
Pulkovo Observatory, Saint-Petersburg, 196140, Russia
         \and 
           Complejo Astron\'omico El Leoncito, Casilla 467, 5400 San Juan, Argentina 
         \and 
Koninklijke Sterrenwacht van Belgi\"e, Ringlaan 3, B-1180 Brussel, Belgium
         \and 
Department of Physics, Astronomy, \& Geology, East Tennessee State University, Johnson City, TN 37614, USA
         \and 
Laboratoire d'Astrophysique, Ecole Polytechnique F\'ed\'erale
de Lausanne (EPFL), Observatoire,
CH-1290~Sauverny, Switzerland
         \and 
Space Sciences Center, University of Southern California, University Park, Los Angeles, CA 90089-1341, USA
          }

\date{October 8, 2007}

\maketitle

\begin{abstract}
We review the recent discoveries of magnetic fields in different types of massive stars and 
briefly discuss strategies for spectropolarimetric observations to be carried out in the future.

\keywords{
stars: abundances --
stars: chemically peculiar --
stars: circumstellar matter --
stars: emission-line, Be --
stars: magnetic fields --
stars: pulsations --
techniques: polarimetric
}
\end{abstract}

%
\section{Introduction}
Massive stars end their evolution, with a final supernova explosion, as neutron stars 
or black holes. The initial masses of these stars range from $\sim$~8--10~M$_\odot$ to 
100~M$_\odot$ or more, which corresponds to spectral types earlier than about 
B2.  While magnetic fields in the sun and solar-like stars have been studied
intensively, very little is known yet about their existence, origin and
role in massive stars.
In spite of considerable indirect evidence only very few direct magnetic field detections 
have been reported so far. Magnetic fields are accessible through the Zeeman effect.
The Zeeman components of spectral lines are polarized and thus permit magnetic fields to be 
measured even in rapidly rotating massive stars where rotation broadening, etc., prevents 
the resolution of Zeeman components.
Currently, direct measurements are achieved only in two O-type stars, 
$\theta^1$\,Ori\,C and HD\,191612 
with longitudinal magnetic field ($\left<B_l\right>$) values of a few hundred Gauss
(Donati et al.\ 2002; 2006),
and in a few early B-type stars. 
In Fig.~1 we demonstrate the excellent potential of FORS\,1 for measuring magnetic fields
in massive  stars.
Our recent FORS\,1 observations with grism 600R of the mean longitudinal 
magnetic field in $\theta^1$\,Ori\,C are compared with the measurements of Wade et al.\ (2006) obtained with the 
MuSiCoS spectrograph. 
This star was the first O-type star with a detected weak magnetic field varying with the rotation 
period of 15.4\,days.
It is obvious that the FORS\,1 measurements are much more accurate showing a smooth sinusoidal curve in spite
of the phase gap between 0.60 and 0.88. 
However, our observations determine a magnetic geometry different from the one deduced by Wade et al.\ (2006).
The maxima and minima of the measured longitudinal field as well as the phases of the field extrema appear 
to be completely different.
Assuming an inclination of the rotation axis to the line-of-sight of $i$=45$^\circ$,
our modeling of the longitudinal field variation constrains the dipole 
magnetic field geometry of  $\theta^1$\,Ori\,C to $B_d\approx900$\,G and $\beta\approx80^\circ$, where 
$B_d$ is the dipole intensity and $\beta$ is the obliquity angle.
In the next sections we present the results of 
our recent surveys of magnetic fields in massive stars carried out with FORS\,1 at the VLT in recent years.

\begin{figure}
\centerline{\includegraphics[width=0.45\textwidth,angle=0,clip=]{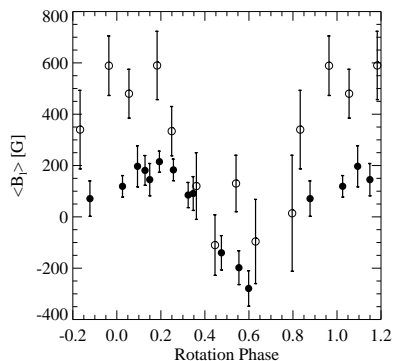}}
\caption{
$\left<B_l\right>$ vs.\ the rotation phase for $\theta^1$\,Ori\,C.
Open circles: Wade et al.\ (2006) with MuSiCoS.
Filled circles: Our FORS\,1 measurements in 2007.
}
\label{fig:1}
\end{figure}

\section{Magnetic fields in SPB and $\beta$ Cephei stars}

We started a systematic search for
magnetic fields in slowly pulsating B (SPB) and $\beta$\,Cephei stars with FORS\,1 in service mode in 2003
(Hubrig et al.\ 2006a). A weak
mean longitudinal magnetic field of the order of a few hundred Gauss was detected
in the $\beta$\,Cephei star $\xi^1$\,CMa and in 13 SPB stars.  The star
$\xi^1$\,CMa became the third magnetic star known among the $\beta$\,Cephei stars. It also 
shows the largest magnetic field and is the hottest magnetic $\beta$\,Cephei star. 
After the publication of these results we obtained two more observing
runs allocated at the VLT. The new observations revealed the presence of magnetic fields in 
ten confirmed SPB and candidate SPB stars and in three $\beta$\,Cephei type stars.
As an example, we present in Fig.~2 the acquired magnetic field measurements of $\xi^1$\,CMa over the 
last 3.7 years. No strong variability or polarity change is detected although a slight increasing trend 
in the strength of the longitudinal field is apparent. 

\begin{figure}
\centerline{\includegraphics[width=0.45\textwidth,angle=0,clip=]{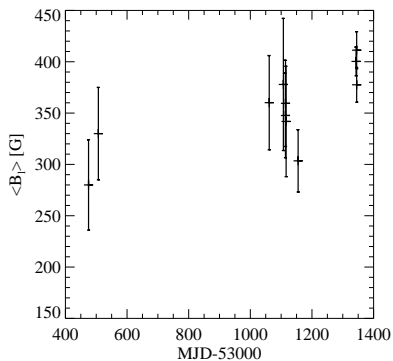}}
\caption{Longitudinal magnetic field measurements of  $\xi^1$\,CMa over the last 3.7 years.
}
\label{fig:2}
\end{figure}

Briquet et al.\ (2007) presented the  evolution of the averaged quadratic effective magnetic 
field $\left<B_l\right>$ in Bp and SPB stars over the main sequence (Fig.~3, left). The value $\log g$ was  used as a 
proxy for the relative age and had the advantage
of being a directly measured quantity.
From this figure it is obvious that the strongest magnetic fields appear in very young Bp stars.
The fact that strong magnetic fields are only observed in a restricted 
range of evolutionary states could be interpreted as a hint for a magnetic field decay in stars at advanced ages.
On the other hand, Hubrig et al.\ (2007a) studied a sample of Ap and Bp stars with accurate Hipparcos 
parallaxes and could show that the magnetic flux remains constant over the stellar life time on the main 
sequence (Fig.~3, right). This result is in full agreement with studies of magnetic fluxes in neutron stars which are
similar to those in magnetic A and B stars and white dwarfs, suggesting that flux conservation during gravitational 
collapse may play an important role (Reisenegger 2007).

\begin{figure}
\begin{center}
\includegraphics[width=0.45\textwidth,angle=0,clip=]{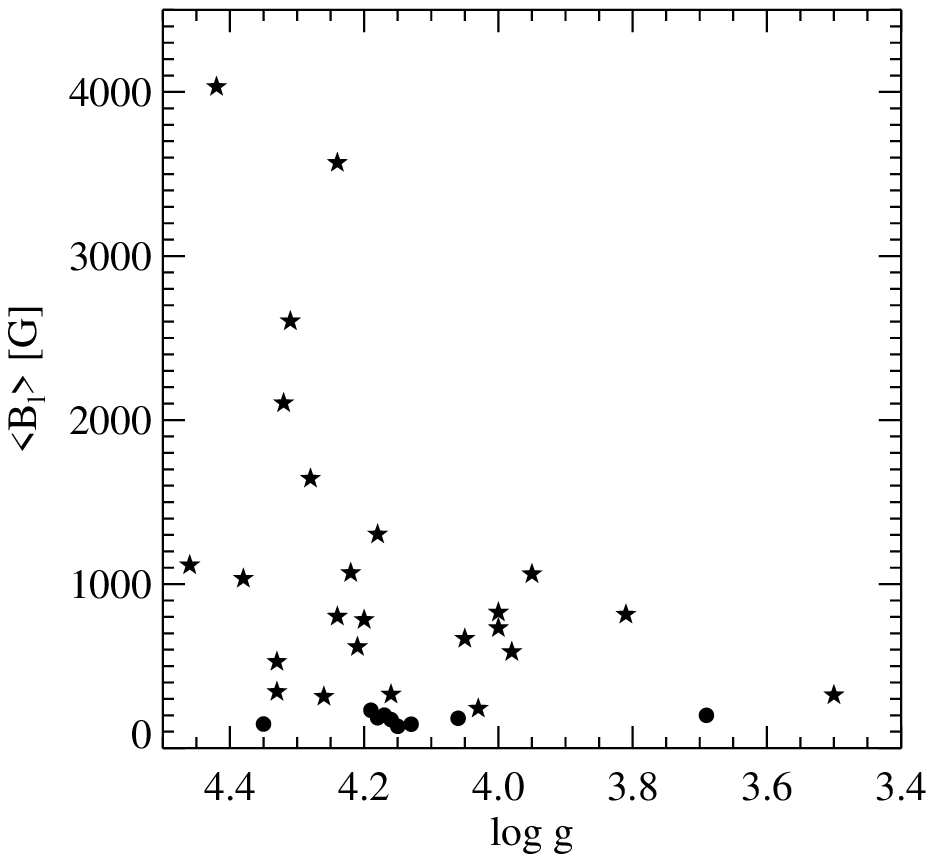}
\includegraphics[width=0.45\textwidth,angle=0,clip=]{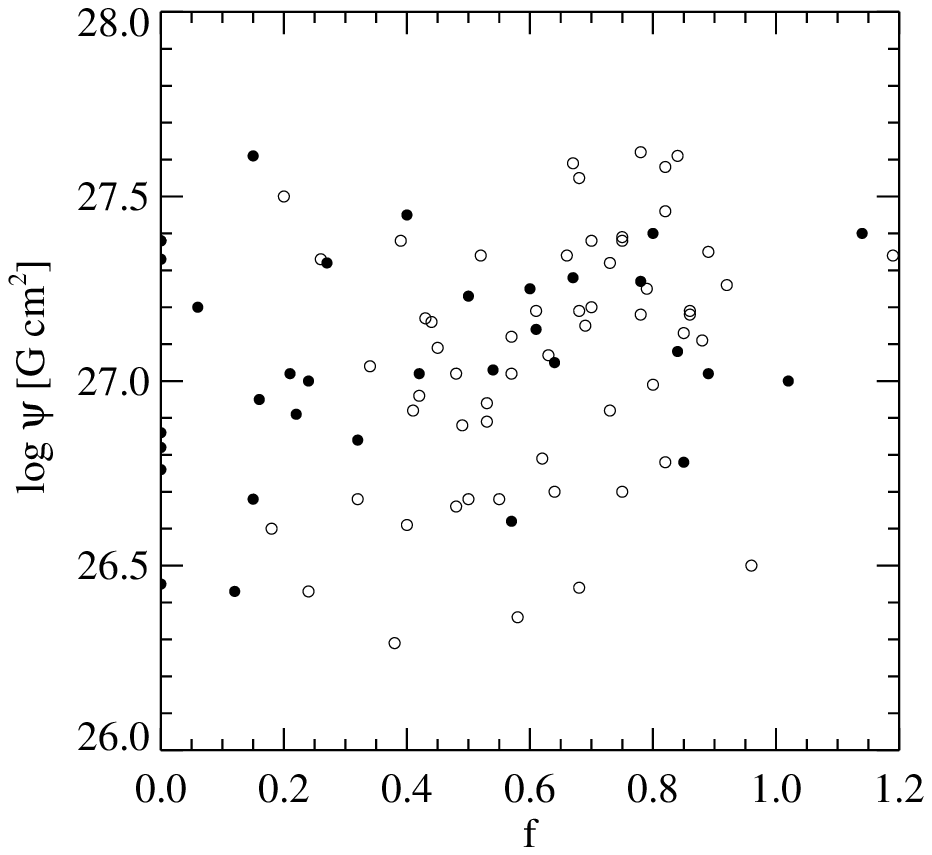}
\end{center}
\caption{
Left: Averaged quadratic effective magnetic field for Bp stars (filled stars) and SPB stars (filled circles)
versus $\log g$.
Right: Magnetic flux in Ap and Bp stars against elapsed time on the main sequence. Filled circles indicate 
stars with mass ${\rm M}>3$\,${\rm M}_\odot$, while open circles indicate stars with mass ${\rm M}<{3}$\,${\rm M}_\odot$.
}
\label{fig:34}
\end{figure}

Very recently, Morel et al.\ (2007) carried out an NLTE abundance study of a
sample of slowly rotating early-type B dwarfs with detected weak magnetic fields. This sample 
includes among other stars also a number of SPB and  $\beta$\,Cephei stars for which we carried out 
the magnetic field survey in recent years.  
The analysis 
strongly supports the existence of a population of nitrogen-rich and boron-depleted slowly rotating 
B stars. The presently available observational data suggest a higher incidence of a nitrogen excess 
in stars with detected magnetic fields.
These results open a new perspective for the selection of
the most promising targets for magnetic field surveys of massive stars using chemical anomalies
as selection criteria.

In summary, our recent observations  of magnetic fields imply that $\beta$\,Cephei stars and SPBs can no 
longer be considered as classes of non-magnetic pulsators. However, the effect of the fields
on the oscillation properties remains to be studied.

\section{Magnetic fields in Be stars}
Be stars are defined as rapidly rotating main sequence stars showing normal O or B-type spectra 
with superposed Balmer emissions.
Until now, weak magnetic fields have been detected in only three Be stars.
A sample of 15~Oe/Be stars was observed with FORS\,1 in April-September 2005 in service mode.
A longitudinal magnetic field at a level larger than 3$\sigma$ has been detected 
in four stars, HD\,56014, HD\,148184, HD\,155806, and HD\,181615 (Hubrig et al.\ 2007b).
Also, an inspection of the Stokes~V spectra of these four stars reveals noticeable Zeeman 
features at the position of numerous spectral lines.
As an example, we present in Fig.~4 the Stokes~I and V spectra for 
HD\,148184 and HD\,155806 in the spectral region around the line He\,{\sc i} $\lambda$ 4471.5\,\AA{}.

\begin{figure}
\begin{center}
\includegraphics[width=0.36\textwidth]{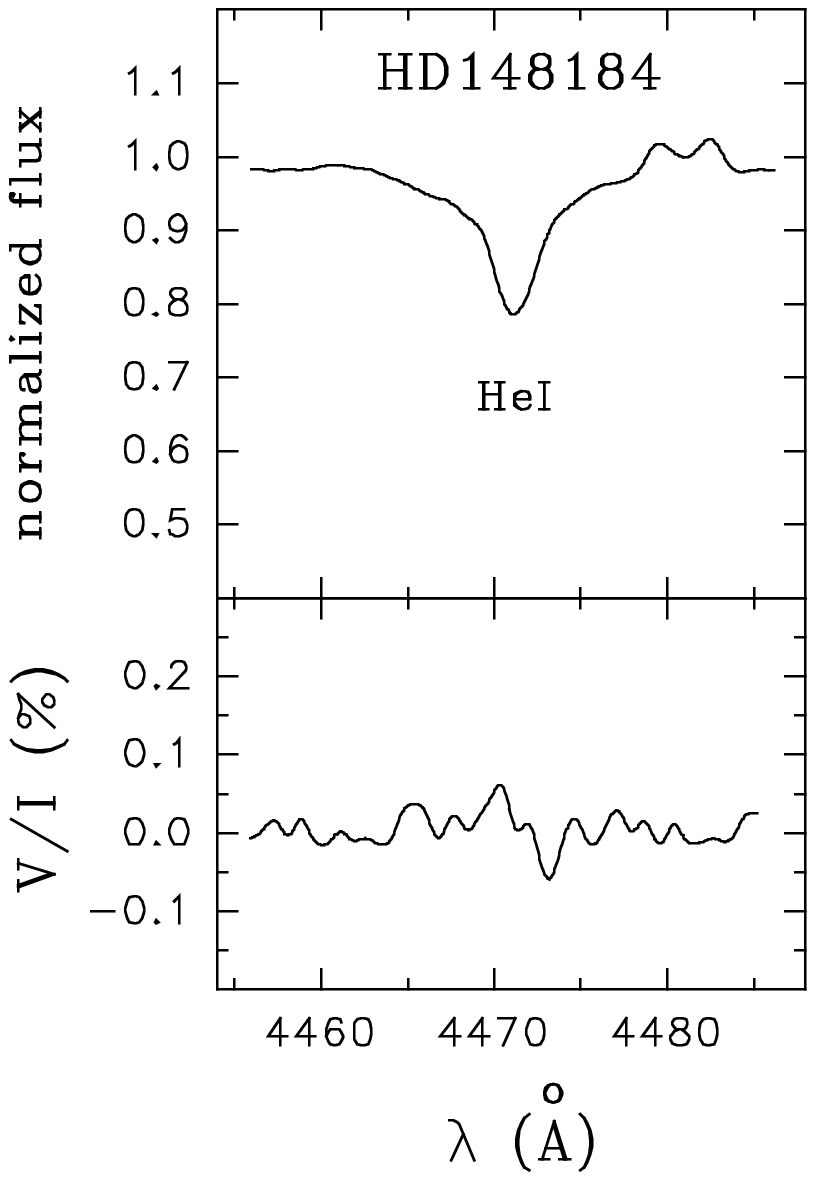}
\includegraphics[width=0.36\textwidth]{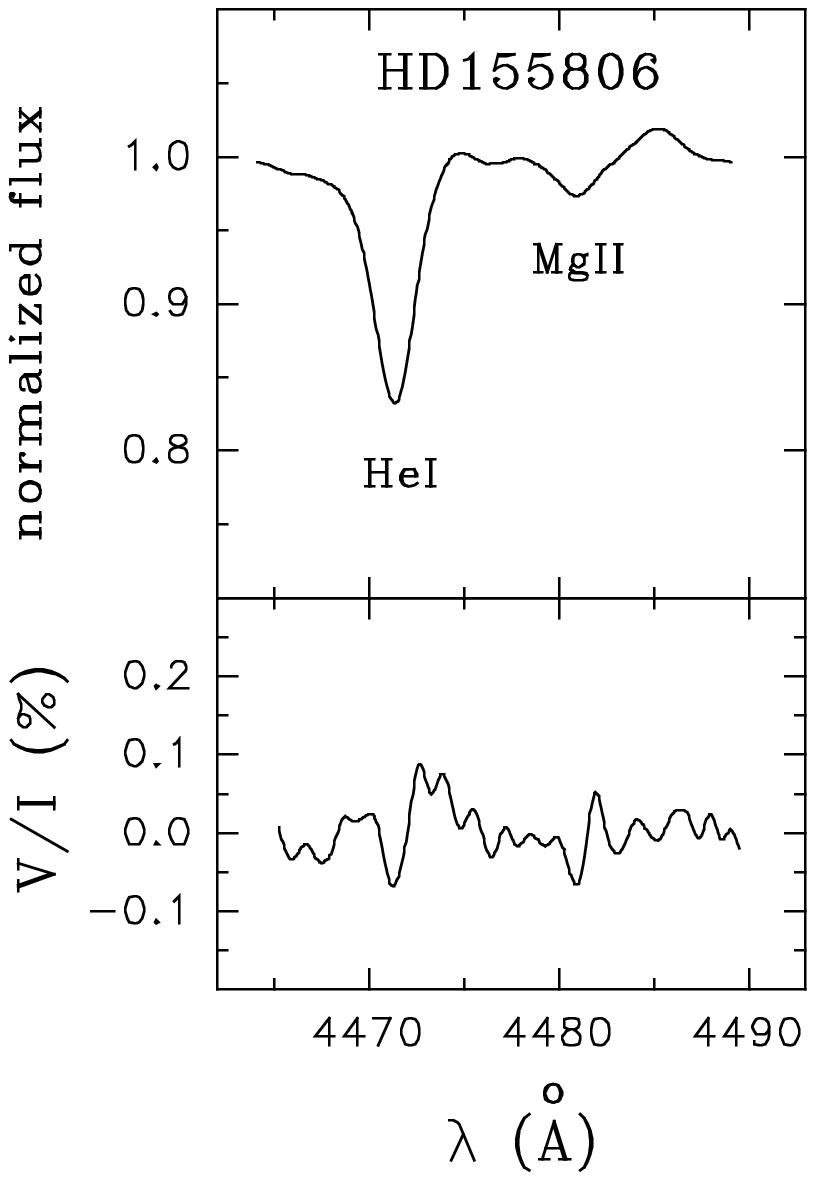}
\end{center}
\caption{Stokes~I and V spectra of HD\,148184 and HD\,155806 in the spectral region around 
the line He\,{\sc i} $\lambda$ 4471.5\,\AA{}.}
\label{fig:5}
\end{figure}

The star HD\,155806 is the hottest star in our sample with a spectral type O7.5IIIe and is currently the 
third O-type star with a magnetic field detected at a level larger than 3\,$\sigma$:
($\langle$$B_z$$\rangle$\,=\,$-$115$\pm$37\,G). Clear variations of Si\,\textsc{iv},\ He\,\textsc{i} and other lines
have been detected in FEROS and UVES spectra retrieved from the ESO archive (Hubrig et al., in preparation). In Fig.\,5
we present the variations of the He\,\textsc{i} 5016\,\AA{} line. 

\begin{figure}
\centerline{\includegraphics[height=0.42\textwidth,angle=270,clip=]{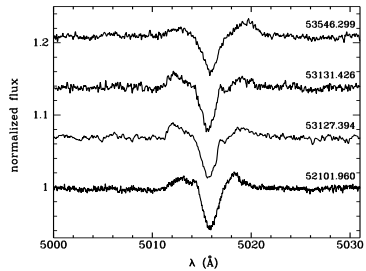}}
\caption{
Spectral profile variability of the He\,\textsc{i} 5016\,\AA{} line in the FEROS and UVES spectra of HD\,155806. 
Spectra are labeled with modified Julian dates.
}
\label{fig:6}
\end{figure}

For three stars in our sample of early type emission stars, HD\,58011, HD\,117357, and HD\,181615,
we noticed the presence of distinctive circular polarization 
signatures detected in the Stokes~V spectra of the Ca\,{\sc ii} H\&K lines. 
The profiles of these Ca lines in the FORS\,1 spectra taken in integral light 
are deeper than predicted by synthetic spectra computed with the  
code SYNTH\,+\,ROTATE developed by Piskunov (1992). Additional high-resolution 
high signal-to-noise spectroscopic observations are needed 
to study the Ca line profiles to be able to decide whether they are formed in the circumstellar disks around 
these stars.
Interestingly, similar types of circumstellar components in 
Ca\,{\sc ii} H\&K lines have recently been discovered by Hubrig et al.\ (2006b, 2007c) in Herbig Ae stars.

\section{Discussion}

Magnetic fields are indeed present in massive stars.
For the case of magnetic fields in non-peculiar massive stars that are weaker in strength
and likely more complex in their geometry than Bp stars,
progress in their study may potentially come from detailed studies of polarized line profiles.
It is not obvious to what extent magnetic fields can be directly discovered in circumstellar material.
Previous detections of magnetic fields in circumstellar material include a detection of magnetic fields in the circumstellar 
disk of FU\,Ori (Donati et al.\ 2005) and in circumstellar Ca lines of Herbig stars 
(Hubrig et al.\ 2006b, 2007c). However,
modeling diagnostics of magnetic fields in these environments are still under development (Ignace \& Gayley 2007).

{}
\end{document}